\documentstyle[prl,aps,floats,epsf]{revtex}
\begin{document}
%\twocolumn
\draft
\title{
{\tenrm\hfill Preprint IFUM -- 467/ FT}\\
Dynamics of N-spins Quantum Ferromagnet with
Uniform Exchange: Exact Solution.}
\author
{Michael Chertkov $^*$ and Igor Kolokolov $^\dagger $}
\address{ $^*$\ Department of Physics of Complex Systems,
Weizmann Institute of Science,
Rehovot 76100, Israel\\ $^\dagger $\ Budker Institute for Nuclear Physics,
Novosibirsk 630 090, Russia \\ and INFN, Sezione di Milano, 20133 Milano, via
 Celoria 16, Italy}
\date{April 8, 1994}
\maketitle
\begin{abstract}
The transverse spin autocorrelation function at an arbitrary temperature
are calculated rigorously for the system of $N$ uniformly exchanged quantum
spins.
At large $N$ the correlator in the para-phase is found to have
a Gaussian bump at small times and a non vanishing tail at large times.
A possible application of those exact results
as forming a starting dynamical mean-field approximation for
long-range quantum magnets and an accompanied physical picture are discussed.
\end{abstract}
\overfullrule=0pt
%\narrowtext
1. Let a system of $N=\sum_j1$ quantum $1/2$-spins be described by the
uniform-range Heisenberg Hamiltonian
\begin{equation}
\hat{\cal H}=-\frac{J}{2N}(\hat{\bbox{\sigma}})^2,
\label{1}
\end{equation}
where $J$ is a value of exchange between spin  and its surrounding  and
$\hat{\bbox{\sigma}}=\sum_j^N \hat{{\bf s}}$ is the total spin
operator of the system.
Suppose for the system to be in thermal equilibrium with an inverse temperature
$\beta$.  Investigation of
static expectation values of the system is
(in the limit $N\gg 1$) the subject of the mean field theory
for magnets with a long-range exchange\cite{Weiss,Vax}.
A question arises: is it possible to create a dynamical mean field approach
for those magnets? Or coming back to the system of $N$ uniformly coupled spins:
what is its equilibrium dynamics?
The uniform feature of the exchange makes the spatial correlations
to be homogeneous what allows to express any spin correlator in terms of
auto-correlation functions.
However, the temporal dynamics of the system turns out
to be extremely complicated; the dynamics of the quantum Heisenberg model
(both in high-temperature \cite{Blum,TerH,Kol1,ChKo} and in low-temperature
\cite{BeLv,Lv,Kol3} phases) remains
interesting if a nonuniform exchange is replaced by the uniform one.

In the present letter, we present for the first time the results of
analytical study of equilibrium dynamics of the system describing
by the Hamiltonian (\ref{1}). A functional integral
formalism used with this model to obtain exact expression for temporal
dependence of the transverse pair auto-correlator at {\em  arbitrary
 temperature
and number of spins}. A physical picture resulting in this evolution is
 suggested.
It is believed that the result is applicable to more
realistic Heisenberg model with a long-range exchange at least as a primary
approximation.

Our starting point is the transverse pair auto-correlator in the following
well-defined form
\begin{equation}
{\cal K}(t,\beta)\equiv\frac{1}{2} \frac{{\bf Tr}[ (\hat{s}^-(0)\hat{s}^+(t)
+\hat{s}^+(0)\hat{s}^-(t))e^{-\beta \hat{\cal H}}]}{{\bf Tr}[e^{-\beta
\hat{\cal H}}]},
\label{2}
\end{equation}
where $\hat{s}^\pm(t)$ are the usual notation for up- and
down- spin operators in
the Heisenberg representation $\hat{s}^\pm(t)=e^{it\hat{\cal H}}\hat{s}^\pm
 e^{-it\hat{\cal H}}$,
$\hat{s}^\pm=\hat{s}^x\pm i\hat{s}^y$.

2.We begin from the infinite temperature when the correlator (\ref{2}) is
\begin{equation}
{\cal K}^0(t)={\bf Tr}[e^{-it\hat{{\cal H}}}\hat{s}^+
e^{it\hat{\cal H}} \hat{s}^-].
\label{3}
\end{equation}
Hubbard-Stratonovich transformations of the evolution operators from
the definition (\ref{3})
%\widetext
\begin{eqnarray}
& & e^{-it\hat{\cal H}}\propto \label{a1a}\int {\cal D}\bbox{\varphi}_1(t)
\exp(-\frac{{\it i}N}{2}\int_0^t dt' \bbox{\varphi}_1^2)\prod_j
{\cal A}^{(1)}_j(t),
\label{4-1}\\
& & e^{it\hat{{\cal H}}}\propto \label{a1b}\int {\cal D}\bbox{\varphi}_2(t)
\exp(\frac{{\it i}N}{2}\int_0^t dt' \bbox{\varphi}_2^2)\prod_j
{\cal A}^{(2)}_j(t),\label{4-2}\\
& & {\cal A}^{(1,2)}_j(t)=Texp({\it i}\int_0^t dt' \bbox{\varphi}_{1,2}(t')
\hat{{\bf s}}_{j})\label{4}
\end{eqnarray}
%%\narrowtext
factorize the initial trace to a product of local ones (here and further we
measure $t$ in the units of $J$).
$T$-exponents in the rhs of (\ref{4-1},\ref{4-2}) may be presented in
terms of new fields $\rho_{1,2},\psi_{1,2}^{\pm}$ as the products
of usual exponents \cite{Kol3}
%\widetext
\begin{eqnarray}
& & {\cal A}^{(1)}(t)=e^{\pi^-_1(t) \hat{ s}^-}
e^{\pi^z_1(t) \hat{s}^z} e^{\pi^+_1(t)
\hat{s}^+},
\label{5-1}\\
& & \pi^-_1(t)=\psi^+_1(t)\ \ , \ \ \psi^+_1(0)=0\ \ , \ \
\pi^z_1(t)=i\int_0^t\rho_1(t')dt',\nonumber\\
& & \pi^+_1(t)=i\int_0^t\psi^-_1(t')\exp(-i\int_0^{t'}
\rho_1(t'')dt'')
dt',\label{5-2}\\
& & {\cal A}^{(2)}(t)=e^{\pi^+_2(t)  \hat{s}^+}e^{\pi^z_2(t)
\hat{s}^z} e^{\pi^-_2(t)
\hat{s}^-},
\label{5-3}\\
& & \pi^{+}_2(t)=\psi^-_2(t)\ \ , \ \ \psi^-_2(0)=0\ \ , \ \
 \pi^z_2(t)=i\int_0^t\rho_2(t')dt',\nonumber\\
& & \pi^-_2(t)=i\int_0^t\psi^+_2(t')\exp(i\int_0^{t'}
\rho_2(t'')dt'')dt',\label{5-4}
\end{eqnarray}
%%\narrowtext
where the new $(\rho,\psi^{\pm})$ and the old $\bbox{\varphi}$
fields are coupled
with each other in the following form
\begin{eqnarray}
 & &\varphi^z_{1,2}=\rho_{1,2}\mp 2\psi^+_{1,2}\psi^-_{1,2},\nonumber\\
&  \varphi^-_1=\psi^-_1 \ \ , \ \
&\varphi^+_1=-i\dot{\psi}^+_1+\rho_1\psi^+_1-(\psi^+_1)^2\psi^-_1,
\nonumber\\
&\varphi^+_2=\psi^+_2
\ \ , \ \
& \varphi^-_2=-i\dot{\psi}^-_2-\rho_2\psi^-_2-(\psi^-_2)^2\psi^+_2.
\label{6}
\end{eqnarray}
The map (\ref{6}) faces the difficulty that it is impossible to express
$T-$exponents in terms of usual function of initial variables $\bbox{\varphi}$;
in general, $\rho$ and $\psi^\pm$ cannot be expressed solely in terms of
${\bf \varphi}$ from (\ref{6}). But if to perform changes of variables
${\bf \varphi}\rightarrow(\rho,\psi^\pm)$ in the functional integrals
(\ref{4-1},\ref{4-2}) it is not necessary to invert (\ref{6}).
The Jacobian ${\cal J}[\rho_{1,2},\varphi^\pm_{1,2}]$:
${\cal D}\bbox{\varphi}_{1,2}={\cal J}[\rho_{1,2},\varphi^\pm_{1,2}]
{\cal D}\rho_{1,2}{\cal D}\psi^\pm_{1,2}$ depends on regularization
of the map (\ref{6}). We make use of the regularization from the works
\cite{Kol3}. It gives for the Jacobian
\begin{equation}
{\cal J}=const \exp(\frac{i}{2}\int_0^t (\rho_1-\rho_2)dt).
\label{7}
\end{equation}
Eqs.(\ref{4-1}-\ref{7}) yield for (\ref{3}), after a calculation of the $N$
local
tracers, the following functional representation for ${\cal K}^0$
\widetext
\begin{eqnarray}
 & &{\cal K}^0=const\int{\cal D}\rho_{1,2}{\cal D}\psi^\pm_{1,2}
e^{{\cal S}_0} {\cal B}^{N-1} {\cal C}
%{\int{\cal D}\rho_{1,2}{\cal D}\psi^\pm_{1,2}
%e^{{\cal S}_0} {\cal B}^N }
\ \ , \ \
{\cal S}_0=-\frac{iN}{2} \int_0^t [\rho_1^2-\rho_2^2-4i\dot{\psi}^+_1\psi^-_1+
4i\psi^+_2\dot{\psi}^-_2]dt+\frac{i}{2}\int_0^t(\rho_1-\rho_2)dt
,\label{8}\\
& &
{\cal B}={\bf Tr}[{\cal A}^{(1)}{\cal A}^{(2)}]=
2\cos[\frac{i}{2}\int_0^t(\rho_1+\rho_2)dt']+
e^{\frac{i}{2}\int_0^t(\rho_1-\rho_2)dt'}
(\psi_1^++i\int_0^t \psi_2^+e^{i\int_0^t\rho_2dt'})
(\psi_2^-+i\int_0^t \psi_1^-e^{-i\int_0^t\rho_1 dt'}),\nonumber
\end{eqnarray}
%\narrowtext
$${\cal C}={\bf Tr}[{\cal A}^{(1)}\hat{s}^+{\cal A}^{(2)}\hat{s}^-]=
e^{\frac{i}{2}\int_0^t(\rho_1-\rho_2)dt'}.$$
Normalization of the functional integral (\ref{8}) depends on $N$ only
and it can be fixed by an evident condition ${\cal K}^0(t=0)=1$.
The transversal fields $\psi^\pm_{1,2}$ have no dynamics at all. Indeed,
the functional integral in (\ref{8}) remains the same if the fields
$\psi^+_2(t'),\psi^-_1(t')$ at an arbitrary moment $0<t'<t$ are replaced by
$\psi^+_2(t),\psi^-_1(t)$ correspondingly.

Our further transformations of (\ref{8})
will be: 1) a lift of ${\cal B}^k$ in the exponential by means of the formula
$$\int_{\Gamma} \frac{dz}{z^{k+1}}e^{-z{\cal B}}=
\frac{\Gamma(k+3/2)}{2\pi i}{\cal B}^k,$$
(here $\Gamma$ is an arbitrary closed around $z=0$ contour in $z$-complex
 plane);
2)Gaussian integrations with respect to $\psi^\pm_{1,2}(t)$;
3)again a lift of a resulting expression in the  exponential, now by
means of the formula
$$ \frac{1}{Y}=\int_0^\infty dy e^{-yY};$$
4)a linear change from dynamical fields $\rho_{1,2}(t')$,$t'<t$ to their
integrals
$\xi_{1,2}(t)=\pm i\int_0^{t'}\rho_{1,2}dt''-
\frac{i}{2}\int_0^t(\rho_1-\rho_2)dt''$.
In total we therefore obtain
\widetext
\begin{equation}
{\cal K}^0(t)=const\int_{\Gamma}\frac{dz}{z^N}\int_0^\infty dy e^{-4N^2y}
\int d\xi_1d\xi_2 \delta(\xi_1+\xi_2) \langle \xi_1|e^{-2z\cosh\xi}
e^{-i\hat{\cal H}_1t}e^{-2\xi}e^{i\hat{\cal H}_1t}|\xi_2\rangle.
\label{9}
\end{equation}
%\narrowtext
Here,
\begin{equation}
\hat{\cal H}_1=-\frac{1}{2N}\partial_\xi^2+2Nzye^{-\xi}.
\label{10}
\end{equation}
The matrix elements in (\ref{9}) with respect to the associated quantum
 mechanics
with the Hamiltonian (\ref{10}) appeared from the Feynmann-Kac path integrals
\cite{FeHi} over ${\cal D}\xi_{1,2}(t)$.
The delta-functions, coupled bra- and ket- states in (\ref{9}),
are established by the boundary condition at the moment $t$,
$\xi_1(t)=-\xi_2(t)$. The initial conditions $\xi_1(0)=\xi_2(0)$ sew together
the direct and inverse
evolution operators in the matrix elements
via its weight of averaging $e^{-2\xi_1(0)}$.
%for the nominator and
%$e^{-\xi_1(0)}$ for the denominator correspondingly).
Let us note, that a similar
matrix element appears at the calculation of the multi-point densities
correlator in $1d$ localization \cite{Kol4}.

Now we can exclude integrations over $y$ in the nominator and denominator of
 (\ref{9}).
It can be done in two steps. First, it is the shift $\xi\rightarrow
 \xi+\ln(4zy)$
and correspondingly
\begin{equation}
\hat{\cal H}_1\rightarrow \hat{\cal H}_2=-\frac{1}{2N}\partial_\xi^2+
\frac{N}{2}e^{-\xi}.
\label{11}
\end{equation}
Second, integration with respect to $y$ itself. Then, equation (\ref{9})
survives if the integration with respect to $y$ and $\exp(-4N^2y)$
with delta-function of $\xi_1+\xi_2$ are replaced
by
$$\frac{F[\eta_1,\eta_2]}{\eta_1\eta_2}=
(\eta_1\eta_2)^{-1}
\exp[-\eta_1\eta_2/(4z)-z(\eta_1^2+\eta_2^2)/(\eta_1\eta_2)],
 $$
where $\eta_{1,2}=2N\exp(-\xi_{1,2}/2)$.
The integral representation for $F[\eta_1,\eta_2]$
$$F=\frac{16}{\pi^2}\int_0^\infty d\nu \nu \sinh 2\pi\nu
K_{2i\nu}(2z)K_{2i\nu}(\eta_1)K_{2i\nu}(\eta_2),$$
with the identity
\[
\eta^{-1}K_{2i\nu}(\eta)=\left(K_{2i\nu+1}(\eta)
-K_{2i\nu-1}(\eta)\right)/4i\nu,
\]
($K_\mu(x)$ it is the usual notation for Macdonald function) yields a further
progress in an evaluation of ${\cal K}^0$.
Indeed, $K_{i\mu}(\eta)$ is an eigen-function of $\hat{\cal H}_2$,
$$\hat{\cal H}_2K_{i\mu}(\eta)=\frac{\mu^2}{8N}K_{i\mu}(\eta).$$
Thus, the matrix elements in (\ref{9}) can be
calculated.
Finally, performing the integrations with respect to $\nu$ and $z$, we obtain
the following answer for the transverse correlator at the infinite temperature
\widetext
\begin{equation}
{\cal K}^0(t)=\frac{2}{3}\{ \frac{1}{2}+(\cos\frac{t}{2N})^N-\frac{1}{N}
((\cos\frac{t}{2N})^{N-2}-1)-(N-1)\sin^2\frac{t}{2N}(\cos\frac{t}{2N})^{N-2}
\}.
\label{12}
\end{equation}
%\narrowtext

${\cal K}^0(t)$ is periodic in time; starting from the unit at zero time it
relaxes to a minimum than restores up
to $1/3$ and  becomes again unit at $t_{per}=4\pi N$.

In the limit of large $N$, when $t_{per}$ is not reached, an
intermediate asymptotic takes place. Thus, at $t=\tau \sqrt{N}, \tau\sim 1,
N\gg 1$
we have a smooth relaxation depending on $N$ via $\tau$ only
\begin{equation}
{\cal K}^0(t)\approx\frac{1}{3}\{
 1+2e^{-\frac{\tau^2}{8}}-\frac{\tau^2}{2}e^{-\frac{\tau^2}{8}}\}.
\label{13}
\end{equation}

The result (\ref{13}) is characterized by a Gaussian bump and a
non vanishing at $\tau\rightarrow\infty$ tail
(see the straight line on the Fig.1).
Let us note, that Gaussian bump has been also obtained by
Belinicher and L'vov for Green function of
dispersion-free magnons in long-range quantum magnets \cite{BeLv}.
The asymptotic value of $K^0(\tau)$ at $\sqrt{N}\ll\tau\ll N$
($1/3$ at $N\rightarrow\infty$) stems from eigenstates of
the initial quantum mechanics (\ref{1}) with the zero full spin  $\sigma=0$.
The non vanishing tail is just an artifact of non-ergodicity of our model.
Thus, in the case of a more realistic long-range model the
$1/3$-plateau can realized in the thermodynamic limit
only as an intermediate time asymptotic.

The behavior (\ref{13}) looks like similar to the result of
calculations \cite{Ku} for spin in a classical random field.
However, in the our case there was no external randomness
at all. And the constant of Gaussian relaxation, had been external
in \cite{Ku}, is defined in our consideration
by means of dynamics itself.

3.  In the recent paper \cite{ChKo} we investigated the long-time dynamics
of an arbitrary-exchange quantum Heisenberg model at the infinite
temperature. It was shown that
the quantum spin pair-correlator (in the para-phase due to unbroken symmetry of
 the Hamiltonian
it is just transverse correlator multiplied by $3/4$) is equal to the
correlator of
a classically
evaluated vector field $\bbox{\phi}_k(t)$
\begin{equation}
\dot{\bbox{\phi}}_k=
\sum_{j}J_{kj} [\bbox{\phi}_k\times \bbox{\phi}_j],
\label{ll}
\end{equation}
averaged over the initial conditions
$\bbox{\phi}_k(0)={\bf p}_k$
with respect to the Gaussian measure
\begin{equation}
\prod_k\it{d}{\bf p}_k\exp\{-\frac{1}{2D}\sum_{i}
{\bf p}_{i}^{2}\}.
\label{llm}
\end{equation}
In the our case $s=1/2$ and $D$ is equal to 1/4.

This "classical" problem, remaining to be strongly nonlinear at an
 arbitrary
exchange $J_{ij}$, becomes linear and exactly solvable for the uniform exchange
$J_{ij}=J/N$. Indeed,  in the uniform case  the right-hand side of (\ref{ll})
 is
$ J[\bbox{\phi}_k\times {\bf P}]/N$, where  ${\bf P}=\sum_k {\bf p}=\sum_k
\bbox{\phi}$
is the integral of motion. Classical motion of a spin turns out to be just
{\em the uniform
precession around the total spin of the system}.
It yields at $N\gg 1, t=\tau \sqrt{N}$ for the transverse auto-correlator
the same answer  (\ref{13}).  Thus, we obtained, first, a good physical picture
resulting in (\ref{13}), second, that transition to the classical model is
valid
 not only at
a large enough time \cite{ChKo}, but also for a long enough exchange rate.

Armed with this understanding of the infinite temperature case we can go
forward
to a finite temperature.

4. It is possible to show that the approach
resulting in (\ref{13}) is generalized on finite temperatures.
Indeed, substitution in the first exponential in the right-hand side of
(\ref{3})
$t_1=t-i\beta$ instead of $t$ produces
${\cal K}^+(t,\beta)$, the real part of which
gives ${\cal K}(t,\beta)$ (\ref{2}).
Thus, eq.(\ref{9}) for ${\cal K}^0(t)$ transforms
into a corresponding one for ${\cal K}^+(t,\beta)$ if $\exp(-i\hat{\cal H}_1t)$
is
 replaced by
$\exp(-i\hat{\cal H}_1t_1)$. Evaluations of the latter equation, performed in
the
 manner
discussed above for the infinite temperature case, results in
\widetext
\begin{eqnarray}
& &{\cal K}^+(t,\beta)=\frac{1}{\sqrt{\beta} Z(\beta)}
\int_{-\infty}^{+\infty} dt'(\cosh t')^N e^{-2t'^2N/\beta}
\{\frac{2Ne^{-t'}}{\beta}(\frac{4t'^2N}{\beta}-t'-1)-\nonumber\\
&
&4\exp(\frac{\beta}{8N}+\frac{2it'}{\beta}(t-i\beta)+\frac{(t-i\beta)^2}{2\beta
 N})
(1+\frac{N}{\beta}-\frac{4t'^2N^2}{\beta^2}+\frac{(4it'N+t-i\beta)(t-i\beta)}
{\beta^2})\},
\label{ful}
\end{eqnarray}
%\narrowtext
where $Z(\beta)$ is the partition function defined, for example,
from this expression with
the condition  $Re[{\cal K}^+(0,\beta)]=1$.

In the intermediate asymptotic $t=\tau\sqrt{N}$, $\tau\sim1$, $N\gg1$
the saddle point approximation of (\ref{ful}) gives a generalization
of (\ref{13}) to the finite temperature case
\begin{equation}
{\cal K}(t,\beta)\approx\frac{1}{3}\{
 1+2(1-\frac{\tau^2}{4-\beta})e^{-\frac{\tau^2}{2(4-\beta)}}\}.
\label{reh}
\end{equation}
Here to avoid huge formulas we restrict ourselves to the para-phase case
$(4-\beta)N\gg 1$ when the only solution of the saddle point equation
\begin{equation}
\tanh t^*=4t^*/\beta,
\label{spe}
\end{equation}
 is $t^*=0$ (in the low temperature case two nonzero solutions
of (\ref{spe}) appear and one
of them has to be chosen as the saddle point in the
corresponding $N\gg 1$ calculations). In this case $\beta<4$,
$Z(\beta)\propto (4-\beta)^{-3/2}$. Thus, we conclude that
at $\beta=4$ there is a peculiarity of usual phase transition type
(of course phase transition exists only in the limit of large
number of spins $N\gg 1$). It is clear that the the static critical exponent
$3/2$ is just mean field (by construction) one but its
evaluation is useful for a control of the really complicated dynamical
calculations.
We see that the squared inverse time
of Gaussian relaxation (\ref{reh}) goes linearly to $\infty$
with $4-\beta \to 0$. The results (\ref{ful},\ref{reh}) are shown graphically
in the Figures.
\begin{figure}[htb]
%\begin{center}
\setlength{\epsfxsize}{65mm}
\leavevmode
\epsffile{fig1.ps}
\caption{Transverse pair auto-correlator as a function
of $\tau$ at $N\rightarrow\infty$ ($\tau=t\protect\sqrt{N}$).
The three plots correspond to different values of inverse temperature:
the straight line - $\beta=0$; the long-dashed
line - $\beta=1$; the dashed line - $\beta=3.5$.}
%\end{center}
\end{figure}
\begin{figure}[htb]
%\begin{center}
\setlength{\epsfxsize}{65mm}
\leavevmode
\epsffile{fig2.ps}
\caption{Transverse pair auto-correlator as a function
of $\tau$ at $\beta=1$. The three plots correspond to different number of
spins:
the straight line - asymptotic at $N\rightarrow\infty$; the long-dashed
line - $N=10$; the dashed line $N=100$.}
%\end{center}
\end{figure}

5. To conclude, we have introduced $N$-spins uniform-exchange quantum model,
for which we have rigorously calculated the temporal dependence of the
transverse
pair auto-correlator at an arbitrary number of spins and temperature.
In the para-phase the correlator shows Gaussian relaxation from
$1$ at $t=0$ via  minimum to
the universal plateau $1/3$ at $t\sim\sqrt{N}\rightarrow\infty$ (Fig.1,2).

The results are obtained by the method that is nothing but
{\em a dynamical mean-field} one and that by the construction turns out to
be {\em exact}. It seems to us that it is the first example of this kind.

{}From the point of view of possible applications to long-range quantum magnets
those exact results are also unique as generating a starting dynamical
approximation
for the para-phase. The possibility to have an exact result for an arbitrary
$N$ is very important. Indeed, the full answer (\ref{ful})
gives an approximation to the problem with large but finite radius of
exchange in an infinite magnet, when $N$ plays the role of a number of spins in
the exchange sphere.
The physical picture accompanied by this approximation is clear from the
classical
model (\ref{ll},\ref{llm}): dynamics of a spin is defined by its uniform
precession
around the full spin of the system.

6. We would like to acknowledge V.Cherepanov, A.Larkin, V.Lebedev and V.L'vov
for their interest in the work and useful discussions.
M.C. is grateful to V.Benza
for the hospitality extended to him during his stay
in INFN, Sez. di Milano,  where a part of this work was done.

\end{document}